\documentclass[12pt]{article}
\usepackage{epsfig}
\usepackage{times}
\date{}

\begin{document}

\title{{\LARGE\sf Short-Range Spin Glasses: 
Results and Speculations}}
\author{
{\bf C. M. Newman}\thanks{Partially supported by the 
National Science Foundation under grant DMS-01-02587.}\\
{\small \tt newman\,@\,cims.nyu.edu}\\
{\small \sl Courant Institute of Mathematical Sciences}\\
{\small \sl New York University}\\
{\small \sl New York, NY 10012, USA}
\and
{\bf D. L. Stein}\thanks{Partially supported by the 
National Science Foundation under grant DMS-01-02541.}\\
{\small \tt dls\,@\,physics.arizona.edu}\\
{\small \sl Depts.\ of Physics and Mathematics}\\
{\small \sl University of Arizona}\\
{\small \sl Tucson, AZ 85721, USA}
}

\maketitle

\begin{abstract}
This paper is divided into two parts.  The first part concerns several
standard scenarios for how short-range spin glasses might behave at low
temperature.  Earlier theorems of the authors are reviewed, and some new
results presented, including a proof that, in a thermodynamic system
exhibiting infinitely many pure states and with the property (such as in
replica-symmetry-breaking scenarios) that mixtures of these states manifest
themselves in large finite volumes, there must be an {\it uncountable\/}
infinity of states.

In the second part of the paper, we offer some conjectures and speculations
on possible unusual scenarios for the low-temperature phase of finite-range
spin glasses in various dimensions.  We include a discussion of the
possibility of a phase transition {\it without\/} broken spin-flip
symmetry, and provide an argument suggesting that in low dimensions such a
possibility may occur.  The argument is based on a new proof of
Fortuin-Kasteleyn random cluster percolation at nonzero temperatures in
dimensions as low as two.  A second speculation considers the possibility,
in analogy to certain phenomena in Anderson localization theory, of a much
stronger type of chaotic temperature dependence than has previously been
discussed: one in which the actual state space structure, and not just the
correlations, vary chaotically with temperature.

\end{abstract}

\vfill\eject

\small
\renewcommand{\baselinestretch}{1.25}
\normalsize

\section{Introduction}
\label{sec:thermolimit}

In this paper we consider Ising spin glass models, in particular the
infinite-range Sherrington-Kirkpatrick~(SK) model~\cite{SK75} and the
nearest neighbor Edwards-Anderson (EA) model~\cite{EA75} on ${\bf Z}^d$.
The SK Hamiltonian (for $N$ spins) is
\begin{equation}
\label{eq:SK}
{\cal H}_N=-(1/\sqrt{N})\sum_{1\le i<j\le N} J_{ij} \sigma_i\sigma_j 
\end{equation}
where the couplings $J_{ij}$ are independent, identically distributed
random variables chosen, e.g.,  from a Gaussian distribution with zero mean and
variance one.

In a series of papers, Parisi and
collaborators~\cite{P79,P83,MPSTV84a,MPSTV84b} proposed, and worked out the
consequences of, an extraordinary {\it ansatz\/} for the nature of this
phase.  Following the mathematical procedures underlying the solution, it
came to be known as {\it replica symmetry breaking\/} (RSB).  The starting
point of the Parisi solution was that the low-temperature spin glass phase
comprised not just a single spin-reversed pair of states, but rather
``infinitely many pure thermodynamic states''~\cite{P83}, not related by
any simple symmetry transformations.

What is primarily of interest is the {\it distribution\/} of the overlaps
between two SK `pure states'~\cite{note} $\alpha$ and $\beta$, defined as
\begin{equation}
\label{eq:qabSK}
q_{\alpha\beta}= {1\over N}\sum_{i=1}^N\langle\sigma_i\rangle_\alpha\langle\sigma_i\rangle_\beta\, ,
\end{equation}
where $\langle\cdot\rangle_\alpha$ is a thermal average in pure state
$\alpha$, and dependence on the coupling realization ${\cal J}$ and
temperature $T$ has been suppressed.  The overlap distribution is
constructed by choosing, at fixed $N$ and $T$, two of the pure states
$\alpha$ and $\beta$ present with probabilites $W_\alpha$ and $W_\beta$ in
the Gibbs state.  The probability that their overlap lies between $q$ and
$q+dq$ is then given by the quantity $P_{\cal J}(q)dq$, where
\begin{equation}
\label{eq:ovdistSK}
P_{\cal J}(q)=\sum_\alpha\sum_\beta W_{\cal J}^{\alpha}W_{\cal J}^{\beta}\delta(q-q_{\alpha\beta})\, .
\end{equation}
As before, we suppress the dependence on $T$ and $N$ for ease of notation.
The average $P(q)$ of $P_{\cal J}(q)$ over the disorder distribution is
commonly referred to as the {\it Parisi overlap distribution\/}, and serves
as an order parameter for the SK model.

The EA Hamiltonian in zero external field is given by
\begin{equation}
\label{eq:EA}
{\cal H}=-\sum_{{{x,y}\atop|x-y|=1}} J_{xy}\sigma_x\sigma_y\ ,
\end{equation}
where the nearest-neighbor couplings $J_{xy}$ are defined in exactly the
same way as the $J_{ij}$ in the SK~Hamiltonian~(\ref{eq:SK}).  In this
model, and unlike in the SK model, thermodynamic pure, mixed, and ground
states are standard, well-defined (see, e.g.,~\cite{NS03jpc,NS02}) objects,
constructed according to well-established prescriptions of statistical
mechanics~\cite{Georgii88,Ruelle,Lanford,Simon,Slawny,Dobrushin,vEvH}.

Now suppose --- as has often been conjectured for finite-range spin
glasses~\cite{BY86,MPV87} --- that there are many pure states not simply
related to each other by symmetry transformations. If this is the case, it
was shown in~\cite{NS92} that local variables, such as correlations, will
vary chaotically and unpredictably as volume size changes (with, say,
periodic boundary conditions on each volume).  Is it then even possible to
describe the nature of large finite volume systems via a single infinite
volume object, and if so, how?  It turns out that an object, called the
{\it metastate\/}~\cite{NS96b} by the authors, can be constructed to
capture the nature of the behavior inside finite systems as volume
increases.  The metastate is a useful tool that describes the empirical
distribution of local variables as volume increases without bound, and is
described in more detail in the companion paper~\cite{NSother} and
in~\cite{NS03jpc,NS96b,NSBerlin,NS97,NS98,AW90}.  The seemingly chaotic
behavior with increasing volume is modelled by random sampling from the
metastate, regarded as a probability measure on the space of Gibbs
states.  Using the metastate approach, we now examine more closely the
nature of the low-temperature spin glass phase.

\section{The trinity of scenarios}
\label{sec:trinity}

There are three scenarios that have received the major share of attention
in the current literature.  Here we refer only to those dealing with the
{\it pure state\/} structure of the low-temperature spin glass phase.
Other pictures have also received a great deal of attention, particularly
the excited state scenario of Krzakala-Martin~\cite{KM00} and
Palassini-Young~\cite{PY00}.  Proving or disproving these or related
pictures is important for achieving a thorough understanding of the
low-temperature physics of spin glasses.  However, because it can be
shown~\cite{NS01,NS02paris} that these pictures describe excitations that
do {\it not\/} alter pure state structure, their presence or absence can be
logically incorporated into any of the three pictures that we are about to
describe (although on heuristic grounds they appear incompatible with the
RSB picture).

Low-temperature pictures of spin glass long-range order and broken symmetry
start with an assumption about the number of pure states ${\cal
N}(\beta,d)$ (which is the same for almost every coupling
realization~${\cal J}$ --- see, e.g.,~\cite{NSBerlin}).  They assume also a
putative critical (inverse) temperature $\beta_c(d)<\infty$ separating a
paramagnetic phase for $\beta<\beta_c$ from a spin glass phase for
$\beta>\beta_c$.  Although there is good
numerical~\cite{BY86,O85,OM85,KY96} and some analytical~\cite{FS90,TH96}
evidence that above some lower critical dimension $d_c^l$ there does exist
such a finite $\beta_c(d)$, there is as yet no proof or even a strong
physical argument supporting such a conjecture.  Moreover, there is no
logical reason why there cannot be {\it two\/} or more phase transitions in
some dimensions.  However, we will not attempt to enumerate all
possibilities here; we will confine ourselves to the most likely scenarios.
We defer to Sec.~\ref{sec:wild} a consideration (mostly for the fun of it)
of some of the more outlandish sounding possibilities.

The actual value of ${\cal N}(\beta,d)$ is not rigorously known at large
$\beta$ for any $d\ge 2$.  There does exist a rigorous
argument~\cite{NS2D00} supporting --- but not completely proving --- the
conjecture that ${\cal N}(\infty,2)=2$, and a heuristic
argument~\cite{NS8D01} supporting the conjecture that ${\cal
N}(\beta<\infty,2<d<8)\le 2$.  It is generally assumed (but also
not proved) that spin-flip symmetry is broken for $\beta_c<\beta<\infty$,
so that pure states come in global spin-flip reversed pairs.

Given all these assumptions, pure state scenarios generally assume either a
single pair of pure states or infinitely many.  Again, there is at this
time no argument proving that one cannot have, say, ${\cal N}(\beta,d)=20$;
it's just difficult to imagine why such a scenario should occur.  There is,
however, a reasonably strong heuristic argument~\cite{NS98} indicating that
if ${\cal N}(\beta,d)=\infty$, it must be an {\it uncountable\/} infinity
(we will prove this in Sec.~\ref{sec:rsbforsrms} for the RSB picture).

While an assumption about ${\cal N}(\beta,d)$ is the {\it starting point\/}
for each of the three pictures we now describe, they are each much more
than a simple assertion about the number of pure states.  In particular,
there are potentially many 2-state or many-state pictures that correspond
to {\it none\/} of these three scenarios.  Here the only aspect of the
three on which we will focus is the relationship among the pure states.

\subsection{Heuristic Description}
\label{subsec:heuristic}

The scenarios are, in order of increasing complexity, the scaling/droplet
(SD) picture~\cite{Mac84,BM85,BM87,FH86,HF87a,FH87b,FH88}, the chaotic
pairs (CP) picture~\cite{NS03jpc,NS02,NS92,NS96b,NS97,NS98}, and the RSB
picture~\cite{P79,P83,MPSTV84a,MPSTV84b,MPV87}.  The first is a 2-state
scenario, while the second and third are many-state scenarios.  The pure
state structure in these pictures is normally described through the Parisi
overlap function~$P(q)$~\cite{MPV87}.  This function needs to be used with
great care in describing pure state structure in short-range models;
see~\cite{NS03jpc,NS98,HF87a} for a discussion of some of the pitfalls and
problems that can occur in its applications.

In the next subsection we provide precise definitions of the pure state
structure of these pictures; we limit ourselves here to brief heuristic
descriptions.  As already noted, scaling/droplet is a two-state picture.
The overlap distribution is therefore simply a pair of $\delta$-functions
at $q=\pm q_{EA}$, where $q_{EA}$ is the Edwards-Anderson order
parameter~\cite{EA75}, regardless of whether replicas and overlaps are
taken before or after the thermodynamic limit is taken, as shown in
Fig.~1~(see below and especially~\cite{NS03jpc,NS02,NSBerlin} for a more
detailed discussion of the difference between the two procedures).

The CP picture considers the possibility of an infinite number of pure
state pairs, but with a trivial overlap structure: all large, finite
volumes would display an overlap structure equivalent to the SD picture
because only a single pair of pure states appears in each of these volumes.
The actual pure state pair, however, varies chaotically as volume changes
because different volumes select different members from the infinite
ensemble of such pairs.  The overlap distribution for the infinite volume
in the CP picture, constructed by choosing replicas and taking their
overlaps {\it after\/} going to the thermodynamic limit, would then
presumably be a simple $\delta$-function at zero overlap (cf.~Fig.~1).

\begin{figure}
\label{fig:sdcp}
\centerline{\epsfig{file=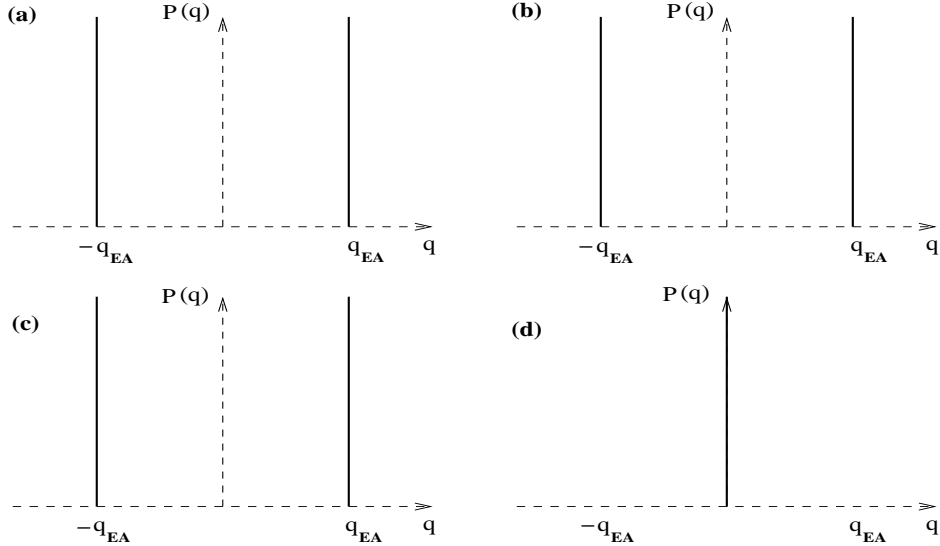,width=3.0in,height=2.5in}}
\caption{
(From Ref.~\cite{NS02}.)
The spin overlap function $P(q)$ at $T<T_c$ for: (a) a two-state
picture when replicas are taken {\it before\/} the thermodynamic limit; (b)
the many-state chaotic pairs picture when replicas are taken before the
thermodynamic limit; (c) a two-state picture when replicas are taken {\it
after\/} the thermodynamic limit; (d) the many-state chaotic pairs picture
when replicas are taken after the thermodynamic limit (conjectured).}  
\end{figure}
\renewcommand{\baselinestretch}{1.25}
\normalsize

There are actually two RSB pictures, which we have called `standard' and
`nonstandard' SK~\cite{NS96b,NS97,NS98,NS96a}.  The first chooses replicas
{\it after\/} taking the thermodynamic limit, and the second before.  Both
have nontrivial overlap structures, which will be described in the next
subsection.

Before turning to that, we need to say a few words about the dimension
dependence of the three pictures.  The only one of these with specific
predictions is the SD picture, which asserts that in {\it every\/} finite
dimension where spin-flip symmetry is broken, there is only a single pure
state pair.  The CP picture does not make any corresponding claim; it
merely asserts that {\it if\/} ${\cal N}(\beta,d)=\infty$ at some
$(\beta,d)$, {\it then\/} the overlap structure must be trivial in the
manner specified above.  If the RSB picture correctly describes the
EA model at $d=\infty$, then both the SD and CP models agree that the upper
critical dimension $d_c^u=\infty$ for the EA model.

The RSB picture is slightly less vague about its dimension-dependence; it
does apparently assume that there exists a strictly finite $d_c^u$ for the
EA model, and that replica symmetry is broken in the nontrivial manner it
specifies for all $d\ge d_c^u$.  The precise value of $d_c^u$
within this picture remains uncertain, but it seems to be higher than two and
less than or equal to three~\cite{FPV94,MPRRZ00}.

\subsection{Description via the Metastate}
\label{subsec:metaversions}

We now turn to a precise description of the three competing pictures, which
is greatly facilitated by using the metastate.

In SD, there is only a single pair of pure states, and these are the same
in every large volume; the overlap distribution function $P_{\cal J}^L(q)$
in a volume $\Lambda_L$ therefore simply approximates a sum of two
$\delta$-functions at $\pm q_{EA}$, as shown in~Fig.~1(a).  In the CP
picture, each finite-volume Gibbs state $\rho^L_{\cal J}$ will still be
approximately a mixture of a {\it single\/} pair of spin-flip-related pure
states, {\it but now the pure state pair will vary chaotically with $L$.\/}
Then for each $\Lambda_L$, $P_{\cal J}^L(q)$ will again approximate a sum
of two $\delta$-functions at $\pm q_{EA}$.

So chaotic pairs resembles the scaling/droplet picture in finite volumes,
but has a very different thermodynamic structure.  It is a many-state
picture, but differs from the RSB picture (see below) in that only a {\it
single\/} pair of spin-reversed pure states $\rho_{\cal J}^{\alpha_L}$,
$\rho_{\cal J}^{\overline{\alpha_L}}$, appears in a large volume
$\Lambda_L$ with symmetric boundary conditions, such as periodic.  In other
words, for large $L$, one finds that
\begin{equation}
\label{eq:possfive}
\rho_{\cal J}^{(L)}\approx {1\over 2}\rho_{\cal J}^{\alpha_L}+{1\over
2}\rho_{\cal J}^{\overline{\alpha_L}}\, ,
\end{equation}
and the pure state pair (of the infinitely many present) appearing in a
particular finite volume $\Lambda_L$ depends chaotically on $L$.  That is,
the periodic b.c.~metastate is dispersed over many 
$\Gamma$'s (in fact, an uncountable
infinity, as we shall see below; this also occurs in the RSB
picture), but (unlike in RSB) each $\Gamma$ is a {\it trivial\/} mixture of
the form $\Gamma = \Gamma^\alpha={1\over 2}\rho_{\cal J}^\alpha+{1\over
2}\rho_{\cal J}^{\overline{\alpha}}$.  The overlap distribution for each
$\Gamma$ is the same: $P_\Gamma = {1\over 2}\delta(q-q_{EA})+{1\over
2}\delta(q+q_{EA})$.  It is interesting to note that there is a spin glass
model (the `highly disordered model'~\cite{NS94,NS96c,BCM94}) that appears
to display just this behavior in its ground state structure above eight
dimensions.

We now turn to the standard and nonstandard mean-field-like scenarios.  The
standard SK picture is perhaps most concisely described in the introduction
of~\cite{FMPP98}.  It requires a Gibbs equilibrium measure $\rho_{\cal
J}(\sigma)$ which is decomposable into many pure states $\rho_{\cal
J}^\alpha (\sigma)$:
\begin{equation}
\label{eq:sum}
\rho_{\cal J}(\sigma)=\sum_\alpha W_{\cal J}^\alpha\rho_{\cal J}^\alpha (\sigma)\ .
\end{equation}

In this picture replicas are taken {\it after\/} going to the thermodynamic
limit.  That is, one chooses $\sigma$ and $\sigma'$ from the product
distribution $\rho_{\cal J}(\sigma)\rho_{\cal J}(\sigma')$, and then the
overlap can be defined as
\begin{equation} 
\label{eq:overlap}
Q=\lim_{L\to\infty}|\Lambda_L|^{-1}\sum_{x\in\Lambda_L}\sigma_x\sigma'_{x}\, ,
\end{equation}
where $|\Lambda_L|$ is the volume of the cube $\Lambda_L$.

Suppose that $\sigma$ is drawn from pure state $\rho_{\cal J}^\alpha$ and
$\sigma'$ from $\rho_{\cal J}^\gamma$.  Then~(\ref{eq:overlap}) equals its
thermal mean~\cite{NS96a}
\begin{equation}
\label{eq:qab}
q_{\cal J}^{\alpha\gamma}=\lim_{L\to\infty}|\Lambda_L|^{-1}
\sum_{x\in\Lambda_L} \langle\sigma_x\rangle_\alpha
\langle\sigma_x\rangle_\gamma \quad ,
\end{equation}
and so the overlap distribution $P_{\cal J}(q)$ is given by
\begin{equation}
\label{eq:PJ(q)}
P_{\cal J}(q)=\sum_{\alpha,\gamma}W_{\cal J}^\alpha W_{\cal J}^\gamma
\delta(q-q_{\cal J}^{\alpha\gamma})\quad .
\end{equation}

According to this picture, the $W_{\cal J}^\alpha$'s and $q_{\cal
J}^{\alpha\gamma}$'s are non-self-averaging quantities, except when
$\alpha=\gamma$ or its global flip, where $q_{\cal J}^{\alpha\gamma}=\pm
q_{EA}$.  The average $P(q)$ of $P_{\cal J}(q)$ over the disorder
distribution $\nu$ of the couplings is then a mixture of two delta-function
components at $\pm q_{EA}$ and a continuous part between them.

There is a technical problem (caused by chaotic size
dependence~\cite{NS92}) in the construction of $\rho_{\cal J}(\sigma)$ that
can be overcome by using the periodic b.c.~metastate $\kappa^{\rm
PBC}_{\cal J}$ (in fact, any coupling-independent metastate would do).  One
can construct~\cite{NS96a} a state $\rho_{\cal J}(\sigma)$ which is the
{\it average\/} over the metastate:
\begin{equation}
\label{eq:rhoav}
\rho_{\cal J}(\sigma)=\int\ \Gamma(\sigma)\kappa_{\cal J}(\Gamma)\
d\Gamma\quad .
\end{equation} 
One can also think of this $\rho_{\cal J}$ as the average thermodynamic
state, $N^{-1}(\rho_{\cal J}^{(L_1)}+ \rho_{\cal
J}^{(L_2)}+\ldots,\rho_{\cal J}^{(L_N)})$, in the limit $N\to\infty$.  It
can be proved \cite{NSBerlin,AW90} that $\rho_{\cal J}(\sigma)$ is a Gibbs
state.

Numerically one constructs overlaps without constructing Gibbs states at
all.  Such a construction (similar to that above) is described in
\cite{NS96a}, and leads ultimately to the same conclusion.

But this picture can be rigorously ruled out for the EA model~\cite{NS96a},
as will be shown in Sec.~\ref{sec:rsbforsrms}.  So the most natural (and
usual) interpretation of a mean-field-like picture cannot be applied to
short-range spin glasses.  The question then becomes: are there
alternative, less straightforward interpretations?  To address this
question, we constructed in~\cite{NS96b} (see
also~\cite{NS03jpc,NS02,NSBerlin,NS97,NS98}) an alternative mean-field-like
picture, the `nonstandard SK model', which we hereafter refer to simply as
the RSB picture. This picture clarifies as well how broken replica symmetry
should be interpreted for the SK model.

Consider again the PBC metastate, although, as always, almost any other
coupling-independent metastate will suffice.  The RSB picture assumes that
in each volume $\Lambda_{L_i}$, the finite-volume Gibbs state $\rho_{{\cal
J},{L_i}}$ is well approximated deep in the interior by a mixed
thermodynamic state $\Gamma^{({L_i})}$, decomposable into many pure states
$\rho_{\alpha_{L_i}}$:
\begin{equation}
\label{eq:gamma}
\Gamma^{({L_i})}=\sum_{\alpha_{L_i}}W_{\Gamma^{(L_i)}}^{\alpha_{L_i}}\rho_{\alpha_{L_i}}\,
\end{equation}
where explicit dependence on ${\cal J}$ is suppressed.  

As in the chaotic pairs picture, the mixed states $\Gamma^{({L_i})}$ change
in some ``chaotic'' fashion with ${L_i}$.  Furthermore, each mixed state
$\Gamma^{(L_i)}$ is presumed to have a nontrivial overlap distribution
\begin{equation}
\label{eq:gamov}
P_{\Gamma^{(L_i)}}=\sum_{\alpha_{L_i},\beta_{L_i}}W_{\Gamma^{(L_i)}}^{\alpha_{L_i}}
W_{\Gamma^{(L_i)}}^{\beta_{L_i}}\delta(q-q_{\alpha_{L_i}\beta_{L_i}})
\end{equation}
of the form shown in Fig.~2.  Moreover, the distances among any three pure
states {\it within a particular\/} $\Gamma$ are assumed to be
ultrametric~\cite{MPV87}.

\begin{figure}
\label{fig:nonstandard}
\centerline{\epsfig{file=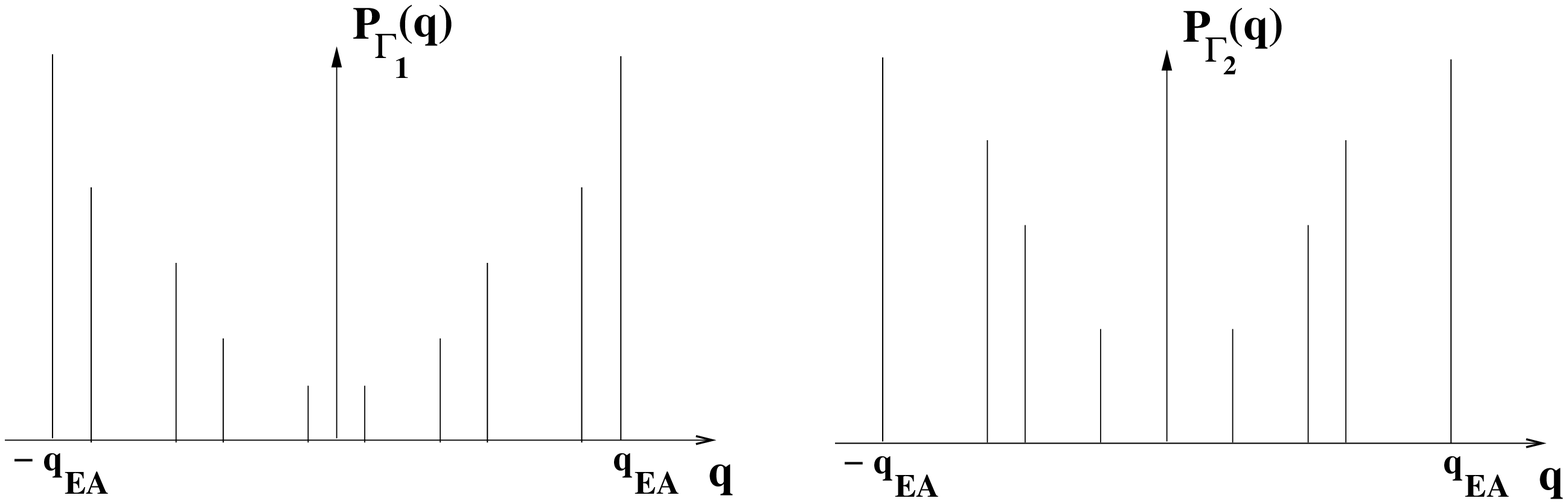,width=7.0in,height=2.5in}}
\caption{(From Ref.~\cite{NS03jpc}.) 
The overlap distribution, at fixed ${\cal J}$, in two different
volumes $\Lambda_1$ and $\Lambda_2$ in the nonstandard SK picture.} 
\end{figure}
\renewcommand{\baselinestretch}{1.25}
\normalsize

We conclude with a brief note about zero temperature.  In each $\Lambda_L$
(with periodic b.c.'s) there can only be a single ground state pair at
$T=0$ (because we are considering Gaussian couplings rather than a $\pm J$
model).  If scaling/droplet holds, then this pair is the same for all large
$L$; if infinitely many ground state pairs exist, then the pair changes
chaotically with $L$. In this respect the behavior of CP is the same at
both zero and nonzero temperatures (below $T_c$), and the same can be said
for SD.  So the overlap functions of CP and RSB differ only at positive
temperature: the mean-field RSB picture at $T>0$ has the $\Gamma^{(L)}$ in
each volume exhibiting a nontrivial mixture of pure state pairs as
in~(\ref{eq:gamma}), while in chaotic pairs the $\Gamma^{(L)}$ appearing in
any $\Lambda_L$ consists of a single pure state pair, as
in~(\ref{eq:possfive}).  There remains, however, an important difference
between the CP and RSB pictures at zero temperature, in that the
space-filling interfaces between ground states should have energies that
scale differently; see~\cite{NS03jpc} and the companion paper in this
volume~\cite{NSother} for a more detailed discussion.

\section{Replica Symmetry Breaking for Short-Range Models}
\label{sec:rsbforsrms}

We begin by eliminating the standard SK picture from further consideration;
the detailed rigorous proof appears in~\cite{NS96a}.  Using the torus-translation
symmetry of the periodic b.c.'s, one can show that the Gibbs state
$\rho_{\cal J}(\sigma)$ is translation-{\it covariant\/}; that is,
$\rho_{{\cal J}^a}(\sigma)=\rho_{\cal J}(\sigma^{-a})$, or in terms of
correlations, $\langle\sigma_x\rangle_{{\cal
J}^a}=\langle\sigma_{x-a}\rangle_{\cal J}$, and similarly for $n$-point
correlations.  Translation covariance of $\rho_{\cal J}$ immediately
implies, via Eqs.~(\ref{eq:overlap})--(\ref{eq:PJ(q)}), translation
invariance of $P_{\cal J}$.  But, given the translation-ergodicity of the
underlying disorder distribution $\nu$, {\it it immediately follows that\/}
$P_{\cal J}(q)$ {\it is self-averaging, and equals its distribution average
$P(q)$ for a.e.~\/}${\cal J}$.  The impossibility of a non-self-averaging
$P_{\cal J}(q)$ can also be shown for other coupling-independent b.c.'s,
where torus-translation symmetry is absent, using methods described in
\cite{NS2D00}.

This leaves nonstandard SK as a maximally allowed mean-field-type picture.
Before discussing its viability, we explain why neither CP nor nonstandard
SK remains viable if modified to have only a {\it countable\/} infinity of
pure states.  We do this first by a heuristic argument valid for both CP
and nonstandard SK, and then by a rigorous argument valid only for
nonstandard SK.  The conclusion of this exercise is that both alternatives
to the SD scenario of a single pure state pair require {\it uncountably
many\/} (i.e., a continuum of) pure state pairs for a single fixed ${\cal
J}$.

Consider $\rho_{\cal J}$, the average over the periodic~b.c.~metastate
$\kappa_{\cal J}$.  Changing from periodic boundary conditions to
antiperiodic, or to any of the many other `partially antiperiodic' boundary
conditions related to periodic ones by a gauge transformation, leaves
$\kappa_{\cal J}$ and consequently $\rho_{\cal J}$ unchanged~\cite{NS98}.
On heuristic grounds, unless $\rho_{\cal J}$ is of the SD form with only a
single pair of pure states, then this lack of dependence on boundary
conditions should hold only if $\rho_{\cal J}$ is a {\it uniform\/} mixture
over infinitely many pure states $\rho_{\cal J}^\alpha$ --- i.e., the
relative weights of all $\rho_{\cal J}^\alpha$'s in $\rho_{\cal J}$ are
equal.  But that is clearly impossible if the number of $\rho_{\cal
J}^\alpha$'s is countably infinite.

In the case of an SK-type picture, where the $\Gamma$'s appearing in the
metastate $\kappa_{\cal J}$ involve nontrivial weights $W_{\cal J}^\alpha$,
one can go further and {\it prove\/} that nontrivial $W_{\cal J}^\alpha$'s
require uncountably many $\rho_{\cal J}^\alpha$'s to appear in
$\kappa_{\cal J}$.  The argument proceeds as follows.  If there are only
countably many $\rho_{\cal J}^\alpha$'s, then the corresponding weights
(say, ordered by magnitude) in $\rho_{\cal J}$ will be measurable and
translation-invariant functions of ${\cal J}$.  By translation-ergodicity,
this means that they are in fact {\it independent\/} of ${\cal J}$ and so
remain the same if finitely many couplings $J_{xy}$ change by amounts
$\Delta J_{xy}$.  Moreover, one can consider the finitely many $\alpha$'s
corresponding to, say, the largest $k$ weights in $\rho_{\cal J}$, and then
the {\it distribution\/} of their weights within the $\Gamma$'s of the
metastate $\kappa_{\cal J}$ is also independent of ${\cal J}$. 

On the other hand, this leads to a contradiction, because a $\Gamma$ in the
nonstandard SK metastate~(see~Eq.~(\ref{eq:gamma})) is a nontrivial
mixture of pure states and their weights.  By the Aizenman-Wehr
transformation~\cite{AW90}, a pure state $\rho_\alpha$ transforms to a pure
state $\rho_{\alpha'}$ under such a finite change ${\cal J}\to{\cal
J}+\Delta J$, and its weight $W_\alpha$ within $\Gamma$ correspondingly
changes according to
\begin{equation}
\label{eq:walpha}
W_\alpha\to W_{\alpha'}=r_\alpha W_\alpha/\sum_\gamma r_\gamma W_\gamma
\end{equation}
where
\begin{equation}
\label{eq:ralpha}
r_\alpha=\langle\exp(\beta\sum_{\langle xy\rangle}\Delta
J_{xy}\sigma_x\sigma_y)\rangle_\alpha\, .
\end{equation}

For pure states $\alpha$ and $\delta$ that are not identical one can find a
choice of $\Delta{\cal J}$, with all $\Delta J_{xy}$ small, such that
$r_\alpha\ne r_\delta$ .  Then for all $\Gamma$'s in which
$W_\alpha(\Gamma)$ and $W_\delta(\Gamma)$ are nonzero, the ratio
$W_{\alpha'}(\Gamma)/W_{\delta'}(\Gamma)$ is changed from
$W_\alpha(\Gamma)/W_\delta(\Gamma)$ by the same factor $r_\alpha/r_\delta$,
that is close to $1$.
But this implies that ${\cal J}\to{\cal J}+\Delta{\cal J}$ will change the
distribution of the weights within the metastate, leading to a
contradiction.  We have thus sketched the proof of:

\medskip

{\bf Theorem 1.}  The PBC metastate $\kappa_{\cal J}$ for the EA spin glass
cannot assign strictly positive probability to $\Gamma$'s whose pure state
decompositions satisfy both:

(i) $\Gamma=\sum_\alpha W_{\cal J}^\alpha\rho_{\cal J}^\alpha$, with not
all $W_{\cal J}^\alpha=1/2$, and

(ii) over all these $\Gamma$'s only countably many $\rho_{\cal J}^\alpha$'s
appear.

\medskip

An immediate consequence of Theorem~1, when combined with our earlier
arguments, is that the only remaining mean-field-like picture is the
nonstandard SK model with a (periodic b.c.)  metastate $\kappa_{\cal J}$
whose average $\rho_{\cal J}$ is supported on an {\it uncountable\/}
infinity of pure states.

We cannot at this point rule this scenario out rigorously, but can provide
a strong heuristic argument, based on a rigorous result, that makes it very
unlikely.  The rigorous result is the already mentioned {\it invariance of
the metastate\/}~\cite{NS98}: two metastates constructed using
gauge-related b.c.'s (e.g., periodic and antiperiodic, or any two randomly
chosen fixed b.c.'s) are {\it identical\/}.  This makes it difficult to see
how any many-state picture can have a $\rho_{\cal J}$ supported on anything
other than a uniform distribution of pure states.  But if this did occur
for a particular ${\cal J}$, an Aizenman-Wehr transformation suggests that
the uniformity would be destroyed for a finitely different ${\cal J}'$.
(For detailed arguments, see~\cite{NS98}.)

Why doesn't this argument also rule out chaotic pairs?  Because in the
chaotic pairs picture, as in scaling/droplet, there are in each
$\Gamma^{(L)}$ only two pure states (although in CP the pair depends on
$L$), each with weight $1/2$.  All even correlations are the same in any
pair of flip-related pure states, so, by Eqs.~(\ref{eq:walpha}) and
(\ref{eq:ralpha}), any change in couplings leaves the weights unchanged.

We conclude that the only viable many-state scenario for short-range Ising
spin glasses is the CP picture with an uncountable infinity of pure states.
This picture has trivial replica symmetry breaking (cf.~Fig.~1) and
consequently a very simple overlap structure.

\section{Wild Possibilities}
\label{sec:wild}

In Sec.~\ref{sec:trinity} we presented the most likely scenarios for the
pure state structure of the spin glass phase in short-range models.  One of
these, the RSB picture, although {\it a priori\/} plausible, especially
given its relevance to the SK model, was rigorously ruled out in two of its
possible versions and excluded heuristically in its final remaining
version.  However, there are other, more exotic scenarios that might also
occur but do not seem to have been considered.  Here, for the sake of
completeness (and perhaps a little whimsy) we suggest two.  In order to
keep the list of possibilities relatively constrained, we will assume in
all that: a) a thermodynamic phase transition does exist above some lower
critical dimension $d_c^l$, and b) in a fixed dimension the low-temperature
phase does not alternate among different scenarios, such as SD and CP, as
temperature is lowered (although the third and most exotic scenario may be
regarded as a kind of phase of extreme alternation).

\subsection{Phase transition without broken spin-flip symmetry}
\label{nobsfs}

As mentioned in Sec.~\ref{sec:trinity}, despite decades of effort and
considerable numerical support~\cite{BY86,O85,OM85,KY96}, there remains no
proof of a phase transition in EA spin glasses.  In this section we provide
a proof that shows that a necessary (though not sufficient) condition for
broken spin-flip symmetry at $T>0$ is satisfied in lattices with dimension
as low as two (e.g., a triangular lattice).  We also discuss a
corresponding sufficient condition.  Of course, this (minimal)
symmetry-breaking is assumed in all scenarios discussed in
Sec.~\ref{sec:trinity}; but here we speculate on the possibility of a phase
transition {\it without\/} the appearance of multiple Gibbs states, and
discuss why the presence of the necessary condition (for broken spin-flip
symmetry) at some $(\beta,d)$ where the sufficient condition is absent
suggests just such a possibility.

Although several approaches are possible, we use here the Fortuin-Kasteleyn
random cluster (RC) representation~\cite{KF69,FK72} extended to
non-ferromagnetic models (see, e.g., \cite{GKN92}), which relates the
statistical mechanics of Ising (or Potts) models to a dependent percolation
problem.  We let ${\bf E}^d$ denote the set of bonds $\langle x,y\rangle$
in some $d$-dimensional lattice ${\bf L}^d$, each with corresponding
coupling $J_{xy}$.  The RC approach introduces parameters $p_{xy}\in[0,1)$
by:
\begin{equation}
\label{eq:pxy}
p_{xy}=1-\exp[-\beta|J_{xy}|]\, .
\end{equation}
The RC distribution is a probability measure $\mu_{\rm RC}$ on
$\{0,1\}^{{\bf E}^d}$, that is, on 0- or 1-valued bond occupation variables
$n_{xy}$.  It is one of two marginal distributions (the other being the
ordinary Gibbs distribution) of a joint distribution on
$\Omega=\{-1,+1\}^{{\bf L}^d}\times\{0,1\}^{{\bf E}^d}$ of the spins and
bonds together, and is given (formally, in the infinite system) by
\begin{equation}
\label{eq:mufk}
\mu_{\rm RC}(n_{xy})=Z_{\rm RC}^{-1}\ 2^{\#(\{n_{xy}\})}\ \mu_{\rm
  ind}(\{n_{xy}\})\ 1_U(\{n_{xy}\})\, ,
\end{equation}
where $Z_{\rm RC}$ is a normalization constant, $\#(\{n_{xy}\})$ is the
number of clusters determined by the realization $\{n_{xy}\}$, $\mu_{\rm
ind}(\{n_{xy}\})$ is the Bernoulli product measure corresponding to
independent occupation variables with $\mu_{\rm ind}(\{n_{xy}=1\})=p_{xy}$,
and $1_U$ is the indicator function on the event $U$ in $\{0,1\}^{{\bf
E}^d}$ that there exists a choice of the spins $\sigma_x$ so that
$J_{xy}n_{xy}\sigma_x\sigma_y\ge 0$ for all $\langle
x,y\rangle$~\cite{SW87,KO88,N93}.  $U$ is the event that there is no
frustration in the occupied bond configuration.  Finite-volume versions of
the above formulas, with specified boundary conditions, are similarly
constructed.

The mapping of this formalism to ferromagnets follows from formulae (that
do not hold for nonferromagnets) such as~\cite{Grimmett99}
\begin{equation}
\label{eq:ferropath}
\langle\sigma_x\sigma_y\rangle=\mu_{\rm RC}(x\leftrightarrow y)\, ,
\end{equation}
where $\langle\sigma_x\sigma_y\rangle$ is the usual Gibbs two-point
correlation function and $\mu_{\rm RC}(x\leftrightarrow y)$ is the RC
probability that $x$ and $y$ are in the same cluster.  Similarly, with
appropriate RC (wired) boundary conditions used for
$\mu_{RC}$, one has 
\begin{equation}
\label{eq:ferroperc}
\langle\sigma_x\rangle_+=\mu_{\rm RC}(x\leftrightarrow\infty)\, .
\end{equation}
It follows that, for ferromagnets, a phase transition from a unique
(paramagnetic) phase at low $\beta$ to multiple infinite volume Gibbs
distributions at large $\beta$ is equivalent to a percolation phase
transition for the corresponding RC measure.

For spin glasses (or other nonferromagnets) the situation is less
straightforward.  Now for two sites $x$ and $y$, (\ref{eq:ferropath})
becomes 
\begin{equation}
\label{eq:sgpath}
\langle\sigma_x\sigma_y\rangle=\langle 1_{x\leftrightarrow\infty}\eta(x,y)\rangle_{RC};
\,\,\,\eta(x,y)=\prod_{\langle x'y'\rangle\in{\cal
C}}{\rm sgn}(J_{x'y'})\, ,
\end{equation}
where ${\cal C}$ is any path of occupied bonds from $x$ to $y$.  By the
definition of $U$, any two paths ${\cal C}$ and ${\cal C'}$ in the {\it
same\/} cluster will satisfy $\prod_{\langle x'y'\rangle\in{\cal C}}{\rm
sgn}(J_{x'y'})=\prod_{\langle x'y'\rangle\in{\cal C'}}{\rm sgn}(J_{x'y'})$.

It is no longer evident that RC percolation is sufficient to prove broken
spin-flip symmetry. Consider the case of a finite volume $\Lambda_L$ with
fixed boundary conditions, i.e., a specification $\overline{\sigma}_x=\pm
1$ for each $\overline{\sigma}_x\in\partial\Lambda_L$.  For the
ferromagnet, by first choosing all $\overline{\sigma}_x=+1$ and then all
$\overline{\sigma}_x=-1$, one can change the sign of the spin $\sigma_0$ at
the origin even as $L\to\infty$.  That is, boundary conditions infinitely
far away affect $\sigma_0$, which is a signature of the existence of
multiple Gibbs distributions.

But for the EA spin glass, it is not clear whether, even in the presence of
RC percolation, there exist any two sets of boundary conditions with the
same effect.  Although the infinite cluster in any one RC realization is
unique, different RC realizations can have different paths from
$0\leftrightarrow\partial\Lambda_L$, leading to different signs for
$\sigma_0$. So percolation 
might still allow for $\langle\sigma_0\rangle \to 0$ as $L\to\infty$.

However, it is easy to see that single RC percolation is at least a {\it
necessary\/} condition for multiple (symmetry-broken) Gibbs phases.  The
contribution to the expectation of $\sigma_0$ from any finite RC cluster is
zero: if a spin configuration $\sigma$ is consistent with a given RC bond
realization within such a cluster, $-\sigma$ is also consistent and equally
likely.  As a consequence, at a $(\beta,d)$ where RC percolation does not
hold, $\langle\sigma_0\rangle=0$ in infinite volume. 
We note that a slightly stronger
version of this argument~\cite{N93} proves that the transition temperature
for an EA $\pm J$ (or other) spin glass, if it exists, is bounded from
above by the transition temperature in the corresponding 
(disordered) ferromagnet.

Is there a modification of this approach that could lead to a proof of
multiple Gibbs phases?  One such possibility is what might be called {\it
double\/} RC percolation.  Here one expands the sample space $\Omega$ to
include two independent copies of the bond occupation variables
(for a given ${\cal J}$ configuration), and defines the variable
$r_{xy}=n_{xy}n'_{xy}$, where $n_{xy}$ and $n'_{xy}$ are taken from the two
copies.  One then replaces percolation of $\{n_{xy}\}$ in the single RC
case with percolation of $\{r_{xy}\}$.  It is not hard to see that this
would be a sufficient condition for the existence of multiple Gibbs phases
(and consequently, for a phase transition).

Single RC percolation for the EA $\pm J$ Ising model in $d>2$~\cite{KO88}
has been proved~\cite{GKN92}.  Here we sketch the outline of a proof
(simpler than the one in~\cite{GKN92}) showing that one has single RC
percolation {\it even in\/} $d=2$ --- e.g., on the triangular lattice.
This is already interesting for the following reason: because we are not
aware of strong numerical evidence of multiple Gibbs states for this
geometry, this may be considered evidence that single RC percolation can
occur in spin glasses without broken spin-flip symmetry.  But there is an
even more interesting potential consequence, which we will discuss below.

The proof uses a standard Fortuin-Kasteleyn-Ginibre (FKG)
argument~\cite{FKG71} using correlation inequalities.  Let ${\bf E}_L$ be
the edge set confined entirely within a volume $\Lambda_L$.  It can be
shown that there exist probability measures $\nu_L$ on $\{0,1\}^{{\bf
E}_L}$ such that, if $f(\{n\})$ and $g(\{n\})$ are nondecreasing real
functions (that is, they do not decrease when any $\{n\}\to\{n'\}$, where
every $n_{xy}=1$ corresponds to $n'_{xy}=1$, but $n_{xy}=0$ can correspond
to either $n'_{xy}=0$ or 1), then they are positively correlated:
\begin{equation}
\label{eq:FKG}
\langle fg\rangle_{\nu_L}\ge\langle f\rangle_{\nu_L}\langle g\rangle_{\nu_L}\, .
\end{equation}

One measure satisfying this property is the independent product
measure~\cite{H60}. Consequently, the marginal distribution (i.e., averaged
over the coupling (signs) in the $\pm J$ model) of satisfied bonds (i.e.,
using the ``satisfaction'' variables $\tilde{n}_{xy} = 1$ (or else $0$) if
$J_{xy} \sigma_x \sigma_y > 0$) at $\beta=0$, satisfies this property.
Operationally, one can think of constructing this set by choosing the spins
{\it first\/} through independent flips of a fair coin, and {\it then\/}
choosing the sign of each bond in the same way.  One then has independent,
density-1/2 bond occupation, which percolates on the triangular
lattice~\cite{SE62, Wi81}.

Consider now the satisfaction variables $\tilde{n}_b$ for $\beta>0$. It is
not hard to show that, at any such fixed $\beta$, and for any bond $b =
\langle xy \rangle$,
\begin{equation}
\label{eq:sat}
P( \tilde{n}_b = 1|\ \{\tilde{n}_{b'} : \ b'\ne b\})\ge 1/2+\epsilon(\beta)\, ,
\end{equation}
where $\epsilon(\beta)$ is a nonnegative function of $\beta$.
Eq.~(\ref{eq:sat}) implies that percolation of
satisfied bonds at finite temperature dominates (in the FKG sense)
percolation at infinite temperature.  Since the RC variables
$n_b$ are obtained from the satisfaction variables $\tilde{n}_b$
by a slight (for large $\beta$) dilution, it follows that at
sufficiently low (but nonzero) temperature one has single RC percolation.

It is presumably the case that on the triangular lattice there is no broken
spin-flip symmetry, and also only a single Gibbs state, at all nonzero
temperatures.  But it is worth entertaining the possibility that single,
but no double RC percolation, {\it does\/} imply some sort of phase
transition {\it but with a single Gibbs state at all nonzero
temperatures\/}.

Let $\beta_c<\infty$ be the inverse RC percolation transition temperature
for the EA model on the triangular lattice.  Consider again the expectation
of the spin at the origin, $\langle\sigma_0\rangle$, in a volume
$\Lambda_L$ with plus boundary conditions.  For $\beta<\beta_c$, the
probability of the site 0 belonging to a cluster reaching the boundary is
bounded from above by $c_0(\beta)e^{-c_1 (\beta)L}$, where each
$c_i(\beta)>0$ is a finite constant.  Therefore $\langle\sigma_0\rangle\le
c_0(\beta)e^{-c_1 (\beta)L}$.

For $\beta>\beta_c$, however, 
the probability that the origin belongs 
to a cluster reaching the boundary is bounded away from
zero as $L\to \infty$.  In
order for no phase transition to take place, there must be at fixed $\beta$
an almost exact cancellation between those RC realizations of `positive'
(in the product of the signs of the couplings) and `negative' paths from
the origin to the boundary.  It is at least conceivable that, while
$\sigma_0\to 0$ as $L\to\infty$, it falls off slower than exponentially ---
perhaps as a power law in $L$.  This would imply a phase transition from a
paramagnetic phase at high temperature to a phase at low
temperature with a {\it unique\/} Gibbs state, 
but one where two-point correlations decay as a power law:
$\langle\sigma_x\sigma_y\rangle \sim |x-y|^{-\alpha(\beta)}$ as
$|x-y|\to\infty$.  This would be analogous to the Kosterlitz-Thouless phase
transition for $2D$ $XY$ models~\cite{KT73}.

\subsection{Highly chaotic temperature dependence}
\label{subsec:really}

We discussed earlier~\cite{NS92} the correspondence between multiple Gibbs
state {\it pairs\/} and the appearance of chaotic size dependence of
correlations in an infinite sequence of volumes with, say, periodic
boundary conditions.  A possibly related phenomenon is the speculated
presence in spin glasses (both infinite- and short-range) of {\it chaotic
temperature dependence\/} (CTD).  Roughly speaking, this refers to an
erratic behavior of correlations, upon changing temperature, on
lengthscales that diverge as the temperature increment goes to zero.  CTD
was predicted~\cite{BM87,FH88,BB87} for the EA spin glass within the SD
context; it also seems to be implied by the RSB theory~\cite{K89,NH93,R94}.
More recent numerical and analytical work (see~\cite{BM99} and references
therein) have led to claims that CTD is not present in either the SK or the
EA model (see also~\cite{BDHV02}), although~\cite{BM02} allows for the
possibility of a weak effect at large lattice sizes.  A perturbative
approach~\cite{RC02} observes a very small effect at ninth order.  At this
time the issue remains unresolved.  Its potential presence in spin glasses
is interesting, however, and represents a qualitatively new thermodynamic
feature of at least some types of disordered systems.

In this section we raise the possibility of a far stronger version of CTD,
which we will call {\it highly\/} chaotic temperature
dependence~(HCTD)~\cite{NS8D01}.  Unlike `ordinary' CTD, where correlations
behave in a chaotic fashion as temperature changes but the global pure
state structure does not, in HCTD the number of pure states in the periodic
b.c.~metastate does not behave in a continuous or even monotonic fashion as
temperature is lowered.  This picture departs radically from any other that
has appeared so far in the literature.

The HCTD scenario is summarized as follows.  As in all other pictures,
there exists a deterministic $T_c$ for a.e.~${\cal J}$, such that for all
$T>T_c$ there is a unique (paramagnetic) Gibbs state.  Suppose one were now
to choose an arbitrary (nonzero) $T<T_c$.  Then with probability one (i.e.,
for almost every ${\cal J}$), there is again a unique infinite volume Gibbs
state (though maybe not with exponential decay of truncated correlations).
Nevertheless, as temperature is lowered from $T_c$ to 0 for a fixed typical
${\cal J}$, there would be a (countably infinite) dense set of
temperatures, {\it depending on ${\cal J}$\/}, with broken symmetric
pure phases for that ${\cal J}$ at those temperatures.

Now this scenario might be dismissed, with some justification, as the
authors' fevered imaginings, but it should be noted that just this sort of
phenomenon does happen in other disordered systems. Consider Anderson
localization in one dimension, with the random Schr\"odinger
operator~\cite{AENSS03}
\begin{equation}
\label{eq:rso}
{\cal H}_\omega = -\nabla^2+V_0(x)+\lambda V_\omega(x)\, ,
\end{equation}
where $\lambda>0$, $V_0(x)$ is a bounded potential periodic in $x$, and
\begin{equation}
\label{eq:rp}
V_\omega(x)=\sum_{n=-\infty}^\infty\eta_n^\omega U(x-x_n)\, .
\end{equation}
In~(\ref{eq:rp}) $n$ runs over the integers, $U(x-x_n)$ is a nonegative
localized potential centered at lattice site $x_n$, and the $\eta_n$ are
i.i.d.~random variables uniformly distributed in $[0,1]$.

Operators of the type given by~(\ref{eq:rso}) are known to have certain
properties -- see, e.g.~\cite{CL90}.  If one first chooses a specific
energy, then for a.e.~realization $V_\omega$ of the random potential, that
energy is part of its continuous spectrum (i.e., is not an eigenvalue of
${\cal H}_\omega$). On the other hand, if one first chooses a specific
realization $V_\omega$, then there is almost surely a (countable) dense set
of eigenvalues (in some appropriate energy interval) of ${\cal H}_\omega$,
and of course this set depends on $\omega$.

Returning to the EA~model in (say) three dimensions, it is amusing to
speculate that a phenomenon like that described above for localization
might occur for EA spin glasses, as follows: a) For every arbitrarily
chosen $T$, there is a unique Gibbs state for a.e.~${\cal J}$; but b) there
exists a (deterministic) $T_c$, such that for a.e.~${\cal J}$, the set of
temperatures $T$ such that there are multiple Gibbs states (e.g., with
$q_{EA}\ne 0$) for that ${\cal J}$ is dense in the entire temperature
interval $[0,T_c]$.  By an application of Fubini's theorem~\cite{Feller},
property a) would necessarily imply that the set of such $T({\cal J})$'s
would have zero Lebesgue measure in the temperature line.

{\it Acknowledgments.}  This research was partially supported by NSF
Grants DMS-01-02587~(CMN) and DMS-01-02541~(DLS).

\renewcommand{\baselinestretch}{1.0}


\small
\begin{thebibliography}{10}



\bibitem{SK75}
D.~Sherrington and S.~Kirkpatrick,
{\em Phys.~Rev.~Lett.\/}~{\bf 35}, 1792 (1975).

\bibitem{EA75}
S.~Edwards and P.W.~Anderson,
{\it J.~Phys.~F\/}~{\bf 5}, 965 (1975).

\bibitem{P79}
G.~Parisi, 
{\em Phys.~Rev.~Lett.\/}~{\bf 43}, 1754 (1979).

\bibitem{P83}
G.~Parisi,
{\em Phys.~Rev.~Lett.\/}~{\bf 50}, 1946 (1983).

\bibitem{MPSTV84a}
M.~M\'ezard, G.~Parisi, N.~Sourlas, G.~Toulouse, and M.~Virasoro, 
{\em Phys.~Rev.~Lett.\/}~{\bf 52}, 1156 (1984). 

\bibitem{MPSTV84b}
M.~M\'ezard, G.~Parisi, N.~Sourlas, G.~Toulouse, and M.~Virasoro, 
{\em J.~Phys.~(Paris)\/}~{\bf 45}, 843 (1984).

\bibitem{note} We gloss over here the serious problems surrounding the
definition of pure states in the SK model.  See the companion paper in
this volume~\cite{NSother} for a more thorough discussion.

\bibitem{NS03jpc}
C.M.~Newman and D.L.~Stein,
{\it J.~Phys.: Condensed Matter\/}~{\bf 15}, R1319 (2003).

\bibitem{NS02}
C.M.~Newman and D.L.~Stein,
{\em J.~Stat.~Phys.\/}~{\bf 106}, 213 (2002).

\bibitem{Georgii88}
H.O.~Georgii, {\em Gibbs Measures and Phase Transitions\/} (de Gruyter
Studies in Mathematics, Berlin, 1988).

\bibitem{Ruelle}
D.~Ruelle, {\em Statistical Mechanics\/} (Benjamin, New York, 1969).

\bibitem{Lanford} 
O.~Lanford, in \newblock {\em Statistical Mechanics
and Mathematical Problems, Lecture Notes in Physics, v.~20\/} (Springer, 
1973), pp.~1-135.

\bibitem{Simon} B.~Simon, {\em The Statistical Mechanics of Lattice Gases v.~1\/},
(Princeton Univ.~Press, Princeton, 1993).

\bibitem{Slawny} 
J.~Slawny, in C.~Domb and J.L.~Lebowitz~(eds.), {\em Phase
Transitions and Critical Phenomena v.~11\/} (Academic Press, London, 1986),
pp.~128-205.

\bibitem{Dobrushin} 
R.L.~Dobrushin and S.B.~Shlosman, in
S.P.~Novikov~(ed.), {\em Soviet Scientific Reviews C.~Math.~Phys.,
v.~5\/} (Harwood Academic Publ.,1985), pp.~53-196.

\bibitem{vEvH}
A.C.D.~van~Enter and J.L.~van~Hemmen, 
{\em Phys.~Rev.~A} {\bf 29}, 355 (1984).

\bibitem{BY86}
K.~Binder and A.P.~Young,
{\it Rev.~Mod.~Phys.\/}~{\bf 58}, 801 (1986).

\bibitem{MPV87}
M.~M\'ezard, G.~Parisi, and M.A.~Virasoro,
{\em Spin Glass Theory and Beyond\/} (World Scientific, Singapore, 1987).

\bibitem{NS92}
C.M.~Newman and D.L.~Stein,
{\it Phys.~Rev.~B\/}~{\bf 46}, 973 (1992).

\bibitem{NS96b}
C.M.~Newman and D.L.~Stein,
{\em Phys.~Rev.~Lett.\/}~{\bf 76}, 4821 (1996).

\bibitem{NSother}
C.M.~Newman and D.L.~Stein,
Local vs.~Global Variables for Spin Glasses, to appear in 2004 Ascona
workshop proceedings.

\bibitem{NSBerlin}
C.M.~Newman and D.L.~Stein, in {\it Mathematics of Spin
Glasses and Neural Networks\/}, ed.~A.~Bovier and
P.~Picco (Birkh\"{a}user, Boston, 1997), pp.~243-287.

\bibitem{NS97}
C.M.~Newman and D.L.~Stein,
{\em Phys.~Rev.~E}~{\bf 55}, 5194 (1997).

\bibitem{NS98}
C.M.~Newman and D.L.~Stein,
{\it Phys.~Rev.~E\/}~{\bf 57}, 1356 (1998).

\bibitem{AW90}
M.~Aizenman and J.~Wehr,
{\it Commun.~Math.~Phys.\/}~{\bf 130}, 489 (1990).

\bibitem{KM00} 
F.~Krzakala and O.C.~Martin, 
{\em Phys.~Rev.~Lett.\/}~{\bf 85}, 3013 (2000). 

\bibitem{PY00}
M.~Palassini and A.P.~Young, 
{\em Phys.~Rev.~Lett.\/}~{\bf 85}, 3017 (2000). 

\bibitem{NS01} 
C.M.~Newman and D.L.~Stein, 
{\em Phys.~Rev.~Lett.\/}~{\bf 87}, 077201 (2001). 
	
\bibitem{NS02paris}
C.M. Newman and D.L. Stein, 
{\it Annales Henri Poincare\'\/}~{\bf 4}, Suppl. 1, S497 (2003).

\bibitem{O85}
A.T.~Ogielski, 
{\em Phys.~Rev.~B\/} {\bf 32}, 7384 (1985).

\bibitem{OM85}
A.T.~Ogielski and I.~Morgenstern, 
{\em Phys.~Rev.~Lett.\/} {\bf 54}, 928 (1985).

\bibitem{KY96} 
N.~Kawashima and A.P.~Young, 
{\em Phys.~Rev.~B\/} {\bf 53}, R484 (1996).

\bibitem{FS90}
M.E.~Fisher and R.R.P.~Singh, 
in {\em Disorder in Physical Systems}, 
edited by G.~Grimmett and D.J.A.~Welsh (Clarendon Press, Oxford, 1990), pp.~87--111.

\bibitem{TH96} 
M.J.~Thill and H.J.~Hilhorst, 
{\em J.~Phys.~I} {\bf 6}, 67 (1996).

\bibitem{NS2D00}
C.M.~Newman and D.L.~Stein,
{\it Phys.~Rev.~Lett.\/}~{\bf 84}, 3966 (2000).

\bibitem{NS8D01}
C.M.~Newman and D.L.~Stein,
{\it Phys.~Rev.~E\/}~{\bf 63}, 16101-1 (2001).

\bibitem{Mac84}
W.L.~McMillan, 
{\em J.~Phys.~C\/}~{\bf 17}, 3179 (1984).

\bibitem{BM85}
A.J.~Bray and M.A.~Moore, 
{\em Phys.~Rev.~B\/}~{\bf 31}, 631 (1985).

\bibitem{BM87}
A.J.~Bray and M.A.~Moore,
{\em Phys.~Rev.~Lett.\/}~{\bf 58}, 57 (1987).

\bibitem {FH86}
D.S.~Fisher and D.A.~Huse, 
{\em Phys.~Rev.~Lett.\/}~{\bf 56}, 1601 (1986). 

\bibitem{HF87a}
D.A.~Huse and D.S.~Fisher,
{\em J.~Phys.~A}~{\bf 20}, L997 (1987).

\bibitem{FH87b}
D.S.~Fisher and D.A.~Huse,
{\em J.~Phys.~A}~{\bf 20}, L1005 (1987).

\bibitem{FH88}
D.S.~Fisher and D.A.~Huse, 
{\em Phys.~Rev.~B\/}~{\bf 38}, 386 (1988).

\bibitem{NS96a}
C.M.~Newman and D.L.~Stein,
{\em Phys.~Rev.~Lett.\/}~{\bf 76}, 515 (1996).

\bibitem{FPV94}
S.~Franz, G.~Parisi, and M.A.~Virasoro,
{\em J.~Phys.~I~(France)}~{\bf 4}, 1657 (1994).

\bibitem{MPRRZ00}
E.~Marinari, G.~Parisi, F.~Ricci-Tersenghi, J.J.~Ruiz-Lorenzo,
and F.~Zuliani, {\em J.~Stat.~Phys.\/}~{\bf 98}, 973 (2000).

\bibitem{NS94}
C.M.~Newman and D.L.~Stein, 
{\em Phys.~Rev.~Lett.\/}~{\bf 72}, 2286 (1994). 

\bibitem{NS96c}
C.M.~Newman and D.L.~Stein, 
{\em J.~Stat.~Phys.\/}~{\bf 82}, 1113 (1996).

\bibitem{BCM94} 
J.R.~Banavar, M.~Cieplak, A.~Maritan, 
{\em Phys.~Rev.~Lett.\/}~{\bf 72}, 2320 (1994).

\bibitem{FMPP98}
S.~Franz, M.~M\'ezard, G.~Parisi, and L.~Peliti,
{\it Phys.~Rev.~Lett.\/}~{\bf 81}, 1758 (1998).


\bibitem{KF69}
P.W.~Kasteleyn and C.M.~Fortuin,
{\it J.~Phys.~Soc.~Jpn.\/}~{\bf 26}, 11 (1969).

\bibitem{FK72}
C.M.~Fortuin and P.W.~Kasteleyn,
{\it Physica\/}~{\bf 57}, 536 (1972).

\bibitem{GKN92}
A.~Gandolfi, M.S.~Keane, and C.M.~Newman,
{\it Prob.~Thy.~Rel.~Fields}~{\bf 92}, 511 (1992).


\bibitem{SW87}
R.H.~Swendsen and J.S.~Wang, 
{\it Phys.~Rev.~Lett.\/}~{\bf 58}, 86 (1987).

\bibitem{KO88}
Y.~Kasai and A.~Okiji,
{\it Prog.~Theor.~Phys.\/}~{\bf 79}, 1080 (1988).

\bibitem{N93} 
C.~Newman, in {\it Probability and Phase Transitions\/},
edited by G.~Grimmett (Kluwer, Dordrecht, 1993), pp.~247-260.

\bibitem{Grimmett99}
G.R.~Grimmett, {\it Percolation} (Springer, Berlin, 1999).

\bibitem{FKG71}
C.M.~Fortuin, P.W.~Kasteleyn, and J.~Ginibre,
{\it Commun.~Math.~Phys.\/}~{bf 22}, 89 (1971).

\bibitem{H60}
T.E.~Harris,
{\it Proc.~Camb.~Phil.~Soc.\/}~{\bf 56}, 13 (1960).

\bibitem{SE62}
M.F.~Sykes and J.W.~Essam,
{\it Phys.~Rev.~Lett.\/} {\bf 10}, 3 (1963).

\bibitem{Wi81}
J.~Wierman,
{\it Adv.~Appl.~Prob.\/} {\bf 13}, 293 (1981).

\bibitem{KT73}
J.M.~Kosterlitz and D.J.~Thouless,
{\it J.~Phys.~C\/} {\bf 6}, 1181 (1973).

\bibitem{BB87}
J.R.~Banavar and A.J.~Bray,
{\it Phys.~Rev.~B\/} {\bf 35}, 8888 (1987). 

\bibitem{K89}
I.~Kondor, 
{\it J.~Phys.~A, Math.~Gen.\/}~{\bf 22}, L163 (1989).

\bibitem{NH93}
M.~Ney-Nifle and H.J.~Hilhorst,
{\it Physica A\/} {\bf 193}, 48 (1993).

\bibitem{R94}
F.~Ritort, 
{\it Phys.~Rev.~B\/} {\bf 50}, 6844 (1994).

\bibitem{BM99} \
A.~Billoire and E.~Marinari, 
{\it J.~Phys.~A, Math.~Gen.\/}~{\bf 33}, L265 (2000). 

\bibitem{BDHV02}
J-P.~Bouchaud, V.~Dupuis, J.~Hammann, and E.~Vincent,
{\it Phys.~Rev.~B\/} {\bf 65}, 024439 (2002). 

\bibitem{BM02} 
A.~Billoire and E.~Marinari, cond-mat/0202473.  

\bibitem{RC02}
T.~Rizzo and A.~Crisanti, 
{\it Phys.~Rev.~Lett.\/}~{\bf 90}, 137201 (2003).

\bibitem{AENSS03}
M.~Aizenman, A.~Elgart, S.~Naboko, J.H.~Schenker, and G.~Stolz,
math-ph/0308023.

\bibitem{CL90}
R.~Carmona and J.~Lacroix,
{\it Spectral Theory of Random Schr\"{o}dinger Operators\/}
(Birkh\"{a}user, Boston, 1990).

\bibitem{Feller}
W.~Feller, {\em An Introduction to Probability Theory and Its
Applications, Vol.~II\/} (Wiley, NY, 1971), p.~124.

\end{thebibliography}
\end{document}